\documentclass[12pt]{article}
\usepackage{epsfig}
\usepackage{amssymb}
\usepackage{amsfonts}

\def\ha{\mbox{$\frac{1}{2}$}}
\def\be{\begin{equation}}
\def\ee{\end{equation}}
\def\ba{\begin{array}{c}}
\def\ea{\end{array}}

\newcommand{\bea}{\begin{eqnarray}}
\newcommand{\eea}{\end{eqnarray}}

\newcommand{\kt}{\rangle}

\begin{document}

\begin{center}

{\Large \bf {

Two patterns of ${\cal PT}-$symmetry breakdown

in a non-numerical
 six-state simulation

 }}


\vspace{10mm}


\vspace{3mm}

 {\bf Miloslav Znojil}

 \vspace{3mm}
Nuclear Physics Institute ASCR, Hlavn\'{\i} 130, 250 68 \v{R}e\v{z},
Czech Republic

{e-mail: znojil@ujf.cas.cz}

\vspace{3mm}



\vspace{5mm}

and

\vspace{5mm}

 {\bf Denis I. Borisov}

 \vspace{3mm}
Institute of Mathematics CS USC RAS, Chernyshevskii str., 112, Ufa,
Russia, 450008

%

and

University of Hradec Kr\'{a}lov\'{e}, Rokitansk\'{e}ho 62, 50003
Hradec Kr\'{a}lov\'{e}, Czech Republic

{e-mail: BorisovDI@yandex.ru }

\vspace{3mm}

\end{center}

\section*{Abstract}

Three-parametric family of non-Hermitian but ${\cal PT}-$symmetric
six-by-six matrix Hamiltonians $H^{(6)}(x,y,z)$ is considered. The
${\cal PT}-$symmetry remains spontaneously unbroken (i.e., the
spectrum of the bound-state energies remains real so that the
unitary-evolution stability of the quantum system in question is
shown guaranteed) in a non-empty domain ${\cal D}^{(physical)}$ of
parameters $x,y,z$. The construction of the exceptional-point (EP)
boundary $\partial{\cal D}^{(physical)}$ of the physical domain is
preformed using an innovative non-numerical
implicit-function-construction strategy. The topology of the
resulting EP boundary of the spontaneous ${\cal PT}-$symmetry
breakdown (i.e., of the physical ``horizon of stability'') is shown
similar to its much more elementary $N=4$ predecessor. Again, it is
shown to consist of two components, viz., of the region of the
quantum phase transitions of the first kind (during which at least
some of the energies become complex) and of the quantum phase
transitions of the second kind (during which some of the level pairs
only cross but remain real).

\section*{keywords}

 quantum theory;
 non-Hermitian observables;
 discrete models of bound states;
 typology of instabilities;

\newpage

\section{Introduction}

In 1998, Bender with Boettcher \cite{BB} conjectured that the
reality of the bound state energy spectra (i.e., the unitarity of
the evolution) might be attributed to the unbroken ${\cal
PT}-$symmetry (i.e., parity times time-reversal symmetry) of the
underlying phenomenological Hamiltonian $H$. Mathematical
formulation as well as implementations of the newly developed theory
were, twelve years later, reviewed and summarized by Mostafazadeh
\cite{ali}. At present it is widely accepted that the manifest
non-Hermiticity of the ${\cal PT}-$symmetric Hamiltonians with real
spectra is fully compatible with the Stone's theorem \cite{Stone}
and with the unitarity of the evolution of the quantum system in
question \cite{book}.

The price to pay for the resolution of the apparent paradox lies in
the necessity of an {\em ad hoc\,} amendment of the Hilbert space.
Simply stated (see, e.g., \cite{SIGMA}), one has to distinguish
between the naively preselected initial, unphysical, ``friendly but
false'' Hilbert space ${\cal H}^{(F)}$ (in which our ${\cal
PT}-$symmetric Hamiltonian with real bound-state spectrum appears
manifestly non-Hermitian,  $H \neq H^\dagger$) and its ``standard
physical'' amendment ${\cal H}^{(S)}$ (here, the inner product is
amended in such a way that the {\em same} operator becomes
self-adjoint,  $H =H^\ddagger$).

The innovative picture of quantum dynamics led to a perceivable
extension of the class of tractable quantum Hamiltonians. For
example, in the traditional unitary quantum theory of textbooks the
linear differential Hamiltonians
 \be
 H = -\triangle + V(\vec{x})
 \label{loca}
 \ee
must be kept self-adjoint in ${\cal H}^{(S)}={\cal H}^{(F)}
=L^2(\mathbb{R}^d)$. In the new context the constraint was relaxed.
The progress was rendered possible by the separation of ${\cal
H}^{(F)} =L^2(\mathbb{R}^d)\neq {\cal H}^{(S)}$. This resulted in
the representation of unitary systems in two different Hilbert
spaces, viz., in physical ${\cal H}^{(S)}$ and, simultaneously, in
auxiliary unphysical ${\cal H}^{(F)}$. A number of innovative
model-building activities followed \cite{Carl}.

Successfully, the mathematical meaning of ${\cal PT}-$symmetry
$H{\cal PT}={\cal PT}H$ was identified with the older concepts of
pseudo-Hermiticity  $H^\dagger{\cal P}={\cal P}H$ \cite{ali}  {\it
alias} Krein-space self-adjointness \cite{Langer} of the
Hamiltonian. Still, for the generic non-Hermitian Hamiltonians the
physical essence of quantum dynamics in ${\cal H}^{(S)}$ appeared
counterintuitive and deeply non-local \cite{Jones}. It has been
revealed that for many non-Hermitian local potentials $V(\vec{x})$
(some of which played the role of benchmark toy models) the amended
physical Hilbert space ${\cal H}^{(S)}$ need not exist, in
mathematical sense, at all \cite{Siegl}.

One of the ways out of the crisis has been found in a return to the
more restricted class of the so called quasi-Hermitian Hamiltonians
$H$. In nuclear physics, for example, these operators were
obligatorily assumed bounded in ${\cal H}^{(F)}$ \cite{Geyer}.
Often, they were even represented by the mere finite,
$N-$-dimensional matrices $H^{(N)}$. In what follows, we shall also
proceed along this line.

In the historical perspective \cite{MZbook} the inspiration of the
latter strategy can be traced back to the Kato's rigorous
mathematical monograph \cite{Kato}. Many illustrative Hamiltonians
were chosen there in the form of matrices with minimal dimension
$N=2$. Also in Ref.~\cite{DDT} devoted to the study of several
manifestly non-Hermitian differential operators (\ref{loca}),
several anomalous spectral features caused by ${\cal PT}-$symmetry
were sucessfully mimicked by certain most elementary benchmark
matrices $H^{(N)}$ with $N=2$.

Due to ${\cal PT}-$symmetry, the bound-state-energy spectrum of
$H^{(N)}$ can be either ``physical'' (i.e., real, compatible with
unitarity) or ``unphysical'' (i.e., containing one or several
non-real, complex conjugate pairs). This leads to one of the most
interesting mathematical questions and challenges in the ${\cal
PT}-$symmetric quantum mechanics: Once we assume that a given ${\cal
PT}-$symmetric Hamiltonian depends on a $J-$plet of couplings or
dynamics-specifying real parameters $g_1=a,g_2=b,\ldots, g_J=z$, we
must be able to separate the Euclidean space $\mathbb{R}^J$ of these
parameters into an  open domain ${\cal D}_{(physical)}^{(N)}$ (in
which our Hamiltonian $H^{(N)}=H^{(N)}(a,b,\ldots, z)$ is
diagonalizable and in which its spectrum is real) and its unphysical
complement (in which the necessary physical Hilbert space ${\cal
H}^{(S)}$ does not exist).

The determination of the domain ${\cal D}_{(physical)}^{(N)}$ and
the localization and description of its boundary  $\partial {\cal
D}_{(physical)}^{(N)}$ are two rather difficult mathematical tasks
in general. The (technically much easier) determination of at least
one of the non-empty subdomains  ${\cal D}_{(0)}^{(N)} \subset {\cal
D}_{(physical)}^{(N)}$ is in fact one of the most important
necessary conditions of the very applicability of the formalism.
This is one of the reasons of the above-mentioned popularity of the
benchmarks with minimal $N=2$. For $N=2$, indeed, the construction
of ${\cal D}_{(physical)}^{(2)}$ proves always feasible by
non-numerical means \cite{nje2}.

The difficulty of the analysis does not grow too much at $N=3$ but
the three-dimensional (and, in general, all odd-dimensional) matrix
models are not too instructive because their ``added'' energy level
remains always real \cite{nje3}. In this context, several existing
studies \cite{nje4,DB} of the next, $N=4$ benchmark matrix spectra
seem to represent a feasibility limit and a transition point between
the numerical and non-numerical descriptions of the ``quantum phase
transition'' boundaries $\partial {\cal D}_{(physical)}^{(N)}$.

In our present paper we intend to develop, in some detail, the idea
as mentioned in Refs.~\cite{maximal,tridiagonal} where it enabled us
to push the non-numerical description of domain ${\cal
D}_{(physical)}^{(N)}$ beyond the $N=4$ boundary. In the language of
mathematics the essence of the idea lies in an {\it ad hoc\,}
lowering of the number $J$ of the variable parameters in matrices
$H^{(N)}(a,b,\ldots, z)$. In the context of physics, a very strong
motivation of such a project may be seen in the fact that the
boundary  $\partial {\cal D}_{(physical)}^{(N)}$ should be, in
general, composed of the hierarchy of the lower-dimensional
subdomains of the parameters at which the quantum system in question
encounters a genuine phase transition.

A complementary, more specific reason for the study of the $N\geq 4$
models has been found in Ref.~\cite{DB}. In a model with $N=4$, the
above-mentioned quantum phase transitions appeared there to be of
{\em two different kinds}. We expect that such an observation should
and could be also reconfirmed at $N=6$. In what follows we will
extend, therefore, some of the qualitative non-numerical $N=4$
results of Ref.~\cite{DB} to the next, phenomenologically richer
model with $N=6$. The model will be introduced in Section
\ref{kapdve} and its properties will be described and discussed in
subsequent Sections \ref{kaptri} and \ref{kapcty}.

\section{Six-state ${\cal PT}-$symmetric model\label{kapdve}}

\subsection{Hamiltonian}

We intend to work with the real matrices $H^{(N)}(a,b,\ldots, z)$
for which the time-reversal ${\cal T}$ is realized by the matrix
transposition and for which the role of the parity-reversal is
played by the antidiagonal $N$ by $N$ square-of-unit matrix
 \be
 {\cal P}= {\cal P}^{(N)}=
  \left[ \begin {array}{ccccc}
 0&0&\ldots&0&1
 \\{}0&\ldots&0&1&0\\
 {}\vdots&
 {\large \bf _. } \cdot {\large \bf ^{^.}}&
 {\large \bf _. } \cdot {\large \bf ^{^.}}
 &
 {\large \bf _. } \cdot {\large \bf ^{^.}}&\vdots
 \\{}0&1&0&\ldots&0
 \\{}1&0&\ldots&0&0
 \end {array} \right]\,
 \label{anago}
 \ee
with the first nontrivial choice of dimension $N=6$. In another
methodical constraint we recall Eq.~(\ref{loca}) and assume that our
toy model Hamiltonian $H^{(N)}$ will be split into an arbitrary
(i.e., non-Hermitian but ${\cal PT}-$symmetric) interaction matrix
$V^{(N)}=V_{ij}^{(N)}$ and a kinetic-energy-simulating term
$\triangle^{(N)}$ given in the form of the standard discrete
Laplacean $\triangle_{ij}^{(N)}=\delta_{i,j+1} + \delta_{i,j-1}$,
i.e., as matrix
 \be
 \triangle^{(N)}=
 \left [ \begin {array}{ccccc}
  0&-1{}&0&\ldots&0
 \\{}-1{}&0&-1&\ddots&\vdots
 \\0&\ddots&\ddots&\ddots&0
  \\{}\vdots&\ddots&-1&0&-1{}
 \\{}0&\ldots&0&-1{}&0
 \end {array}
 \right ]\,.
 \label{kinuu}
 \ee
Although the general form of matrix $V^{(N)}=V_{ij}^{(N)}$ could
contain, in principle, up to $J_{max}=N^2$ (i.e., in the present
paper, $J_{max}=36$) free real parameters, we shall follow the
guidance by paper \cite{DB} and keep the matrix form of the real
interaction term strictly tridiagonal and antisymmetric. Together
with the requirement of ${\cal PT}-$symmetry this makes our ultimate
$N=6$ toy model Hamiltonian strictly $J-$parametric with $J=N/2=3$,
 \be
 H^{(6)}(x,y,z)=
 \left [
 \begin{array}{cccccc}
 0&-1+z&0&0&0&0\\
 -1-z&0&-1+y&0&0&0\\
 0&-1-y&0&-1+x&0&0\\
 0&0&-1-x&0&-1+y&0\\
 0&0&0&-1-y&0&-1+z\\
 0&0&0&0&-1-z&0
 \ea
 \right ]
 \,.
 \label{se16ch}
 \ee

\subsection{Bound-state energies}

The general discrete Schr\"{o}dinger equations for bound states
 \be
  H^{(N)}\,|\psi^{(N)}_n\kt= E^{(N)}_n\,|\psi^{(N)}_n\kt\,,
  \ \ \ \ \ \ n = 0, 1, \ldots, N-1\,,\ \ \ \ N<\infty\,
  \label{se}
  \ee
may be solved, in practice, by various computer-assisted numerical
methods. Under the present choice of Hamiltonian (\ref{se16ch}) with
$N=6$, fortunately, the related spectrum-determining secular
equation
 \be
 \det
 \left [
 H^{(N)}(x,y,z)-E^{(N)}_n\,I^{(N)}
 \right ]
 =0\,
 \label{se16ch}
 \ee
admits a perceivable simplification mediated by the change of
variables
 \be
 z \to C=1-z^2\,,\ \ y
 \to B=1-y^2\,,\ \ x \to A=1-x^2\,.
 \label{16ch}
 \ee
After some algebra one converts the secular equation into polynomial
relation
 \be
 {{\it E}}^{6}+ \left [ -2\,C-2\,B-A \right ] {{\it E}}^{4}+ \left [ 2\,
 BC+2\,AC+{C}^{2}+{B}^{2} \right ] {{\it E}}^{2}-A{C}^{2}=0\,.
 \label{e19}
 \ee
The equation is solvable in terms of Cardano formulae. This
indicates the non-numerical origin of the pictures which sampled the
parametric-dependence of the energies $E_n^{(6)}(C,B,A)$ in
Ref.~\cite{DB}. Unfortunately, the shape of domain ${\cal
D}_{(physical)}$ as well as the specification of its boundaries
$\partial {\cal D}_{(physical)}$ were omitted there as tractable by
the purely numerical constructive means.

The general numerical algorithm of the latter construction was
described (but not numerically tested) in Ref.~\cite{horizon}. Thus,
the conclusions of the two studies \cite{DB,horizon} were
discouraging: the construction of the boundary $\partial {\cal
D}_{(physical)}$ remains a purely numerical task at $N=6$. Moreover,
even a sufficiently transparent presentation of the results of the
numerical construction seem to require an active use of some
interactive graphical software.

In what follows we intend to demonstrate that the sceptical
conclusions of Ref.~\cite{horizon} resulted from the consideraation
of too broad a class of matrices $H^{(6)}$. In this context, the
main result of our present study will lie in the discovery that due
to certain specific features of our choice of Hamiltonians
(\ref{e19}), the construction of the physical unitary evolution
domain ${\cal D}_{(physical)}$ may be made much more
straightforward. We shall show that up to the necessary
determination of certain auxiliary constants, this construction also
remains strictly non-numerical at $N=6$.

\section{Domain ${\cal D}_{(physical)}^{(6)}(A,B,C)$\label{kaptri}}

From the matrix form of our present $N=6$ Hamiltonian (\ref{se16ch})
one can immediately deduce that in a way generalizing the results of
Ref.~\cite{DB} the boundary $\partial {\cal D}_{(physical)}$ will
contain, {\it i.a.}, the three planes $A=0$, $B=0$ and $C=0$
representing the Kato's \cite{Kato} exceptional points (EPs) at
which the matrix ceases to be diagonalizable.

Such an overall reality-domain-structure conjecture will be given a
more complete, constructive and explicit form in what follows. It is
also time for us to point out now that our present approach will be
based on the replacement of the tedious and practically useless
explicit Cardano formulae for $E_n(C,B,A)$ by their perceivably
simpler and analytically tractable implicit-function alternatives.

In this setting, one of the other and also one of the most important
properties of the set ${\cal D}_{(physical)}$ will be guessed (i.e.,
conjectured and, later, proved) via an analogy with the $N=4$
results of Ref.~\cite{DB}. This property is that the admissible
innermost coupling $A$ must be always positive.

\subsection{An elimination of the outermost coupling
$C=1-z^2$}

In a preparatory step let us replace the (positive) parameter $A$ by
a pair of its real square roots $\alpha=\pm \sqrt{A}$ (by our
above-mentioned conjecture, these roots remain always real). Via the
subsequent application of the computer-assisted factorization
techniques (followed by their not too difficult backward check) we
managed to reduce our secular Eq.~(\ref{e19}) to the pair of
formulae
 \be
 C=C^{(\pm)}(E,B,A)=E^2-\frac{B}{E-\alpha^{(\pm)}(A)}\,E\,
 \label{formuc}
 \ee
for the outermost coupling (one for each sign of $\alpha$). Such an
$N=6$ analogue of its $N=4$ predecessor exhibits a parabolic
asymptotic growth $C^{(\pm)}(E,B,A)=E^2 +{\cal O}(1)$ at $|E| \gg
1$. One of the generic zeros of $C=C(E)$ occurs at
$E=E_{zero}^{(0)}=0$. The resulting estimated shape is further
modified by a single first-order pole at (positive or negative)
$E_{singular}=\alpha^{(\pm)}(A)$. This means that at the large $C$s
and for each sign of $\alpha \lessgtr 0$ the curve will be
intersected by the horizontal line $C={\rm const.}$ at the three
different real values of the bound state energy $E_n$. Thus, after
one (numerically) determines the appropriate EP minimum
$C_{(EP)}=C_{(EP)}(B,A)$ of the function $C=C(E)$, all of the points
$C \in ( C_{(EP)}(B,A),\infty)$ (and only these points) will belong
to ${\cal D}_{(physical)}$.

Without loss of generality (i.e., due to left-right symmetry of the
spectrum of energies) let us now consider just the branch of
Eq.~(\ref{formuc}) with positive $\alpha>0$ (i.e., with the pole to
the right from the origin). We then can distinguish between the two
alternative scenarios in which the remaining two zeros of the curve
$C(E)$, viz., values
 \be
 E_{zero}^{(\pm)} = \ha \left (\alpha \pm \sqrt{\alpha^2+4B}
 \right)
 \label{zeroes}
 \ee
are separated by the origin (for $B>0$) or not (for $B<0$; remember
that the two respective limits with $B= 0$ become unphysical).

In the former case with $B>0$ and to the right from $E=\alpha$ the
real intersection of the curve $C(E)$ with the horizontal line
$C={\rm const.}$ will exist at any $C \in (-\infty,\infty)$,
defining the largest (and always real) energy root. The other two
bound state energies will be smaller than $\alpha$, originating from
the intersection of the fixed horizontal line $C>C_{(EP)}(B,A)$ with
the left, U-shaped part of the curve (\ref{formuc}). This curve will
attain its unique minimum (equal to $C_{(EP)}(B,A)$) at the negative
value of energy $E_{min}$. This constant may be evaluated using its
implicit-function definition
 \be
 {\alpha}B=-{2E_{min}}(E_{min}-\alpha)^2\,,
 \ \ \ \ \ B>0\,,
 \ \ \ \ \ E_{min}<0\,.
 \label{extremecy}
 \ee
The other, more interesting scenario occurs at the negative values
of $B<0$. To the right from $E=\alpha$, the curve $C(E)$ will then
become U-shaped, with a real minimum
$C_{right}=C_{(EP)}(B,A)>\alpha^2$. This minimum is attained at the
largest root $E_{right}>\alpha$ of Eq.~(\ref{extremecy}) and it
moves slowly upwards with the decrease of $B < 0$.

At the small negative $B$s and to the left from the pole at
$E=\alpha$, the curve $C(E)$ decreases, with the growth of $E$ from
its zero at $E=0$ to its local minimum $C_{(min)} $ at $E_{min}$. It
subsequently grows to its local maximum  $C_{(max)} $ at $E_{max}$
and, finally, it decreases to minus infinity at $E=\alpha$. The
values $E_{min}$ and  $E_{max}$ are the two remaining roots of the
cubic Eq.~(\ref{extremecy}). In and only in the interval of
(negative) $B \in (B_{(EP)},0)$ these roots remain real. In such an
interval, the domain ${\cal D}_{(physical)}$ also contains an
additional, anomalous, non-empty interval of the acceptable
couplings $C \in ( C_{(min)}(B,A),C_{(max)}(B,A))$, separated from
the above-mentioned upper interval by a non-empty gap of unphysical
$C \in (C_{(max)}(B,A)),C_{(EP)}(B,A))$.

One can conclude that the existence of the gap of a non-zero width
in ${\cal D}_{(physical)}$ should be perceived as a
phenomenologically highly interesting consequence of the
simultaneous variability of the three parameters in our six-site
discrete ${\cal PT}-$symmetric quantum lattice.

\subsection{The innermost  coupling $A=1-x^2$}

Our constructive considerations of the preceding paragraph were
based on the assumption that the domain  ${\cal D}_{(physical)}$
does not contain a part with the negative values of the innermost
parameter $A$. At $N=6$ the validity of this property appears as
surprising and serendipitous as at $N=4$. Also its proof is again
easy to verify only after a nontrivial preliminary computer-assisted
factorization of certain components of the secular polynomial.

The net result of these algebraic manipulations may be expressed by
the following elementary formula
 \be
 \pm\alpha^{(\pm)}(A)=\alpha(E,C,B)=\left (1-\frac{B}{E^2-C}\right )\,E\,.
 \label{formua}
 \ee
Its form looks very similar to its $C-$related predecessor so that
also the analysis of the shape of this function is feasible and
routine.

The analysis is more straightforward at the positive $C=\gamma^2$
when it leads to the simplification
 \be
 \alpha(E,C,B)=E-
 \frac{B}{2}\,
 \left (
 \frac{1}{E+\gamma} +
 \frac{1}{E-\gamma}
 \right )\,.
 \label{33}
 \ee
At the positive $B$s this function of $E$ is composed of the three
separate branches, each of which grows from $-\infty$ to $+\infty$.
As long as the sextuplet of the bound state energies $E_n$ is
specified by the intersections of this curve with the pair of
horizontal lines $\alpha=\pm \sqrt{A} = const.$, the conclusion is
that all of these energies are real iff $A \geq 0$.

After we move to the negative $B$s, the branches of $\alpha(E)$ will
all flip and decrease near the singularities (i.e., at $E \approx
\pm \gamma$). One can only guarantee the entirely robust existence
of a single maximum at $\alpha=\alpha_{(max)}<0$ for
$E=E_{max}<-\gamma$ and of a single minimum $\alpha_{(min)}>0$ at
$E_{min}>\gamma>0$ (this minimum may be found sampled in the right
upper corner of Fig.~\ref{fixu}). Thus, the robust reality of the
sextuplet of the bound state energies can be safely guaranteed in
the interval of the sufficiently large $ A\in
(\alpha_{(max)}^2,\infty)$.

%
\begin{figure}[h]                     
\begin{center}                         
\epsfig{file=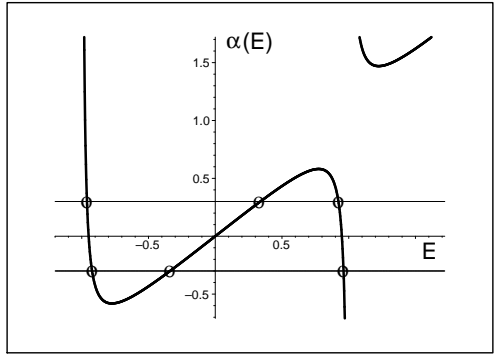,angle=270,width=0.4\textwidth}
\end{center}                         
\vspace{-2mm} \caption{The graph of function $\alpha(E)$ of
Eq.~(\ref{33}) at $B=B_0=1/10$ and $C=C_0=1$ (i.e., $\gamma=1$). The
two thin horizontal lines sample the choice of $A=A_0=0.09$ (i.e.,
of $\alpha_0^{(\pm)}=\pm \sqrt{A_0}=\pm 0.3$). The intersections (=
small circles) determine the six real bound-state energies
$E=E_n(A_0,B_0,C_0)$ with $n=0,1,\ldots,5$.
 \label{fixu}}
\end{figure}

As long as the negative $B$s remain small, $B \in (B_{(EP)},0)$, the
curve $\alpha(E)$ may also exhibit a negative minimum
$\alpha_{(-)}<0$ at some $E_{-} \in (-\gamma,0)$, followed by a
positive maximum $\alpha_{(+)}=-\alpha_{(-)}  >0$ at some $E_{+}=-
E_{-} $ (this situation is sampled in Fig.~\ref{fixu}). One may
conclude that domain ${\cal D}_{(physical)}$ will contain another
component of $A \in (0,\alpha_{(+)}^2)$. This will be an anomalous
subdomain separated from the upper bulk part by the non-empty gap of
unphysical values of $A \in (\alpha_{(+)}^2,\alpha_{(max)}^2)$.

In the last step of our analysis we have to move to the range of
negative $C$s. In this case the correction to the linear term is a
bounded function of $E$ so that the function $\alpha(E)$ itself
(having no real singularities) can only develop a negative local
minimum $\alpha_{left}<0$ at a negative energy (plus, symmetrically,
a positive local maximum $-\alpha_{left}>0$ at an opposite positive
energy). This can only happen at some sufficiently large
$B>B_{(EP)}>0$. In such a case one notices the emergence of a new
piece of domain ${\cal D}_{(physical)}$ with $A \in
(0,\alpha_{left}^2)$.

\subsection{The
intermediate coupling $B=1-y^2$}

For the completion of our description of the geometric shape of
${\cal D}_{(physical)}$ is is necessary to derive and recall also
the last pair of the eligible explicit definitions
 \be
 B^{(\pm)}(E,C,A)=(E-\alpha^{(\pm)}(A))\,(E^2-C)/E\,.
 \label{formub}
 \ee
Such a function with the single singularity in the origin (with the
most elementary growing or decreasing shape $B(E) = \alpha C/E-C +
{\cal O}(E)$) and with the parabolic asymptotic behavior $B(E) =
E^2+ {\cal O}(E)$  always contains a U-shaped part (to the right
from the origin for the positive product $\alpha C$ and {\it vice
versa}). The other part decreases or, respectively, increases from
infinity to infinity so that at least two energy levels are always
real. At the positive $C$s the minimum of the U-shaped part is zero
so that in ${\cal D}_{(physical)}$ we have all $B\in (0,\infty)$. At
the negative $C$s the  minimum $B_{(EP)}$ of the U-shaped part of
the curve $B(E)$ is positive so that the spectral reality constraint
implies that $B \in (B_{(EP)},\infty)$.

\section{Discussion\label{kapcty}}

\subsection{Stable quantum systems in non-Hermitian representations}

Quantum systems exhibiting ${\cal PT}$ symmetry were made popular,
by Bender and Boettcher \cite{BB}, via a one-parametric family of
non-Hermitian quantum Hamiltonians $H=H(\lambda) \neq
H^\dagger(\lambda)$ such that $H(\lambda){\cal PT}={\cal
PT}H(\lambda)$. The spectrum was shown real and discrete if and only
if $0<\lambda < \infty$, i.e., if and only if the value of the
parameter belonged to an ``admissible'' open set ${\cal
D}_{(physical)}$. The authors conjectured that whenever $\lambda \in
{\cal D}_{(physical)}$, these Hamiltonians may be given the
conventional unitary-evolution-generator interpretation.

The expectations were confirmed. Several reviews
\cite{ali,book,Carl} may be consulted for a detailed account of the
well-developed quantum theory covering ${\cal PT}$ symmetric quantum
systems. In the Bender's and Boettcher's toy model, in particular,
the physical domain of parameters ${\cal
D}_{(physical)}^{(BB)}=(0,\infty)$ is a semi-infinite interval. As a
consequence the decrease of $\lambda$ and its passage through the
boundary $\lambda^{(BB)}=0$ results in the loss of the reality of
the spectrum and, simultaneously, in the breakdown of the ${\cal
PT}$ symmetry of the system. This breakdown is spontaneous, i.e.,
the wave functions suddenly lose the symmetry while the Hamiltonian
itself remains formally ${\cal PT}$ symmetric.

The knowledge of the boundaries of the physical domain of parameters
becomes important when one turns attention to the study of
mechanisms of the loss of quantum stability. An exactly solvable
illustrative example may be provided by the non-self-adjoint but
${\cal PT}$-symmetric harmonic-oscillator Hamiltonian
 \be
 H^{(HO)}(\lambda)=-\frac{d^2}{dx^2}
 +\frac{\lambda^2-1/4}{(x - {\rm i}\varepsilon)^2}
 +(x - {\rm i}\varepsilon)^2\,,
 \ \ \ \ x \in (-\infty,\infty)
 \label{diffee}
 \ee
in which the shift $\varepsilon\neq 0$ is arbitrary~\cite{ptho}. In
this model the parameter $\lambda=0$ is still the point of
spontaneous breakdown of ${\cal PT}$-symmetry. Below this value the
energy spectrum ceases to be real. Nevertheless, the physical domain
of parameters cannot contain any positive integers (see the proof in
\cite{ptho}) so that it becomes ``punctured'' and topologically
nontrivial,
  \be
  {\cal D}_{(physical)}^{(HO)}=\left \{
  \lambda \in (0,\infty)\,, \lambda \notin \{1,2,\ldots\} =\mathbb{Z}^+
  \right \}\,.
  \label{hops}
  \ee
The excluded integer parameters have a number of interesting
properties. Although the energies merge at these EP singularities,
they {\em do not complexify\,} in their vicinity. The operator
$H^{(HO)}(\lambda_{(EP)})$ itself ceases to be diagonalizable and it
acquires a Jordan-block canonical form at these values. Naturally
\cite{ali}, such an operator does not admit any acceptable
probabilistic physical interpretation.

\subsection{Exactly solvable models: phenomenological
appeal\label{nonhermiti}}

The exact solvability of quantum systems is a vague concept. It
covers the most elementary harmonic oscillators as well as certain
truly complicated many-body systems characterized by sophisticated
symmetries. In between the two extremes one finds a number of
systems called quantum lattices or $N-$site quantum chains, the
dynamics of which is controlled by a finite-dimensional
Schr\"{o}dinger equation (\ref{se}). Indeed, the choice of a finite
matrix dimension $N<\infty$ makes these models solvable,
numerically, with arbitrary precision.

Recently, the practical studies of the quantum models of this type
found a challenging new motivation in the context of the growth of
interest in the non-Hermitian quantum Hamiltonians with real spectra
\cite{MZbook}. Some of the puzzling obstacles posed by rigorous
mathematics were successfully circumvented by the restriction of
attention either to the ${\cal PT}-$symmetric Hamiltonians
\cite{Carl} or to the class of bounded operators \cite{Geyer} or, in
an extreme case, to the finite-dimensional matrices as sampled in
Eq.~(\ref{se}).

On phenomenological side people managed to connect some of these
branches of the theory with experiments. In the laboratory they
confirmed various predicted properties of $N-$point-lattice
structures characterized by the ${\cal PT}-$symmetric balance
between gain (sources) and loss (sinks) \cite{ring1}. One of the
most interesting as well as challenging features of all of the
classical as well as quantum non-Hermitian lattices may be seen in
the possibility of having the latter symmetry, at certain values of
parameters, spontaneously broken \cite{regions}. In the most common
scenario this breakdown means that some eigenenergies complexify and
survive as complex conjugate pairs (cf., e.g., the implementation of
the idea of the possible complexification in the context of
non-Hermitian quantum thermodynamics \cite{tight}).

\subsection{Quantum physics near the real exceptional points}

For couple of years the problem of the passage of $\lambda$ through
the level-crossing points $\lambda_{(EP)}^{(k)}\in \mathbb{Z}^+$ of
model (\ref{diffee}) remained unclarified. The changes of physics at
the level-crossing EP boundaries $\partial {\cal D}^{}$ were known
to be model-dependent and technically difficult to describe
\cite{Batal,JHC}. Many researches came to the conclusion that by
non-numerical means, the explicit constructive explanation of the
underlying physics may not even be feasible at all. One of the key
sources of the scepticism lied in the complicated mathematics.
Dieudonn\'{e} \cite{Dieudonne}, for example, discouraged the
applications of the non-Hermitian operators in physics rather
persuasively.

The resolution of the  contradictions was provided by the
quasi-Hermitian quantum theory \cite{ali,SIGMA,Geyer,last}. In the
framework of this theory it has been clarified that the energies may
remain real even if the Hamiltonian itself appears non-selfadjoint.
It became widely known that this operator can be made self-adjoint
via an {\it ad hoc} change of the physical inner product in Hilbert
space. Thus, whenever our parameters stay inside the physical domain
${\cal D}$, many non-Hermitian Hamiltonians $H(\lambda)$ may be
assigned the status of an acceptable quantum observable.

One of the most important keys to the applicability of the theory
lies in the correct localization of the physical domains $ {\cal
D}^{(HO)}$. Naturally, the theory ceases to be applicable in the
limit $\lambda \to \lambda_{(EP)}$. In the language of physics, the
{\em observable aspects} of the quantum system in question may
change whenever its {real} variable parameter $\lambda$ {\em
crosses\,} the singular dynamical boundary at $\lambda_{(EP)}$.

In the related literature the attention is almost exclusively payed
to the scenarios in which the loss of the observability involves the
Hamiltonian. In the language of Ref.~\cite{DB} one can speak about
the quantum phase transition of the first kind. Typically, a new
degree of freedom emerges and must necessarily be included in an
amended Hamiltonian. In model (\ref{diffee}) the latter change only
takes place at the leftmost point
$\lambda_{(EP)}=\lambda_{(EP)}^{(0)}=0$ of the boundary  $\partial
{\cal D}^{(HO)}$ at which all of the energies cease to be real. In
contrast, the passage of the parameter through any other EP value
$\lambda_{(EP)}=\lambda_{(EP)}^{(k)}=k$, $k = 1, 2, \ldots$ does not
lead to any complexification. The observability of the energy
survives and we may hold the Hamiltonian unaltered.

\subsection{Matrix models: mathematical
appeal\label{nomiti}}

The advantage of solvability of the extremely elementary $N=2$
matrix models is already playing the role of an inspiring methodical
guide for years \cite{ali,Kato,shendr}. Unfortunately, one has to
pay for the advantage. The facilitated mathematical tractability of
the low-dimensional models may be more than counterbalanced by their
perceivably smaller phenomenological appeal. One also has to mention
their limited capability of leading to a deeper insight or of
enhancing the predictive power of quantum theory.

Using the toy models it has been pointed out~\cite{Denis,Denisa}
that the quantum lattices can support not only the well known
spontaneous breakdown of ${\cal PT}-$symmetry but also another,
alternative version of phase transition, {\em not accompanied} by
the complexification of the energies. In review \cite{MZbook}, for
example, it was emphasized that both of the complexifying and
non-complexifying quantum phase transitions have the same
mathematical origin and that both of them may be attributed to the
non-Hermiticity of at least one of the observables. This opened a
new viable model-building perspective in which one of the Kato's
complex points of degeneracy loses its characteristic imaginary part
and becomes {\em real}.

In place of the realistic but difficult differential-operator
Hamiltonians sampled by Eq.~(\ref{diffee}) we recommended, in
Ref.~\cite{DB},  the use of certain elementary $N=4$ predecessor of
our present $N=6$ model. Re-written in the form
 \be
 H^{(4)}(\lambda)=\left [ \begin {array}{cccc}
 0&-1+\sqrt{1-\lambda}&0&0\\\noalign{\medskip}-1-
 \sqrt{1-\lambda}&0&-1+\sqrt{1-A}&0
 \\\noalign{\medskip}0&-1-\sqrt{1-A}&0&-1+\sqrt{1-\lambda}
 \\\noalign{\medskip}0&0&-1-\sqrt{1-\lambda}&0\end {array} \right ]\,
  \,,
 \label{wright4}
 \ee
the latter Hamiltonian was found observable for $\lambda$ in
  $$
  {\cal D}_{(physical)}^{(4)}=
  (-A/4,0)\bigcup (0,\infty)\,
  $$
i.e., inside an elementary but still ``punctured'' analogue of the
harmonic-oscillator physical domain (\ref{hops}). After such a
simplification of mathematics an important progress was achieved in
physics because we were able to prove that the new system's passage
through the unavoided-level-crossing point $\lambda_{(EP)}^{(4)}=0$
{\em does\,} change the system of observables (other than
Hamiltonian). One can certainly speak about the quantum phase
transition of the second kind.

In Ref.~\cite{DB} we did not manage to make our argumentation
sufficiently model-independent. For the purely technical reasons the
extension of our matrix model from $N=4$ to $N=6$ was temporarily
found too difficult. In this sense we just filled the gap in our
present paper. We showed that the main phenomenological observations
about the coexistence of the quantum phase transitions of the first
and second kind in a single quantum system may be expected to be
generic.


\subsection*{Acknowledgements}

M.Z.~was supported by GA\v{C}R Grant Nr.~16-22945S and by IRP
RVO61389005. D.B.~was partially supported by grant of RFBR, by the
grant of President of Russia for young scientists-doctors of
sciences (MD-183.2014.1) and by Dynasty fellowship for young Russian
mathematicians.

\end{document}